\documentstyle[aps,pre,eqsecnum]{revtex}

\def\eps{\varepsilon}
\def\Dm{\widetilde{\cal D}_{\mu}}
\def\E{\overline\epsilon}

\begin{document}
\draft

\title{RENORMALIZATION GROUP IN THE STATISTICAL THEORY OF TURBULENCE:
TWO-LOOP APPROXIMATION}
\author{L.Ts.~Adzhemyan, N.V.~Antonov, M.V.~Kompaniets, A.N.~Vasil'ev}
\address{Department of Theoretical Physics, St~Petersburg University,
Uljanovskaja 1, St~Petersburg, Petrodvorez, 198504 Russia}

\maketitle
\begin{abstract}
The field theoretic renormalization group is applied to the
stochastic Navier--Stokes equation that describes fully developed fluid
turbulence. The complete two-loop calculation of the renormalization
constant, the beta function and the fixed point is performed.
The ultraviolet correction exponent, the Kolmogorov constant and the
inertial-range skewness factor are derived to second order of the
$\varepsilon$ expansion. Submitted to {\bf Acta Physica Slovaca}.
\end{abstract}

\section{Introduction}

One of the oldest open problems in theoretical physics is that of describing
fully developed turbulence on the basis of a microscopic model. The latter
is usually taken to be the stochastic Navier--Stokes (NS) equation
subject to an external random force which mimics the energy input by the
large-scale modes; see, e.g., \cite{Monin,Legacy}. The aim of the theory is
to verify the basic principles of the celebrated Kolmogorov--Obukhov
phenomenological theory, study deviations from this theory, determine
the dependence of various correlation functions on the times, distances,
external (integral) and internal (viscous) turbulence scales, and derive
the corresponding scaling dimensions. Most results of this kind were obtained
within the framework of numerous semiphenomenological models which cannot be
considered to be the basis for construction of a regular expansion in certain
small (at least formal) parameter \cite{Monin,Legacy}.

An important exception is provided by the renormalization group (RG) method
that was earlier successfully applied in the theory of critical behaviour
to explain the origin of critical scaling and to calculate universal
quantities (critical dimensions and scaling functions) in the form of the
$\eps$ expansions \cite{Zinn}.

The RG approach to the stochastic NS equation, pioneered in [4--7], allows
one to prove the existence of the infrared (IR) scale invariance with
exactly known ``Kol\-mo\-go\-rov'' dimensions and the independence of the
correlation functions of the viscous scale (the second Kolmogorov
hypothesis), and calculate a number of represen\-ta\-ti\-ve constants in
a reasonable agreement with experiment. Detailed review of the RG theory
of turbulence and more references can be found in \cite{UFN,turbo}.

In contrast to the standard $\phi^{4}$ model of critical behaviour
\cite{Zinn}, where the critical exponents are known up to the order
$\eps^{5}$ (five-loop approximation), all the calculations in the RG
approach to the stochastic NS equation have been confined with the simplest
one-loop approximation. The reason for this distinction is twofold.
First, the multiloop calculations for this dynamical model are rather
involved: one can say that the two-loop calculation for the stochastic NS
equation is as cumbersome as the four-loop calculation for the conventional
$\phi^{4}$ model. Second, the critical dimensions for the most important
physical quantities (velocity and its powers, frequency, energy dissipation
rate and so on) are given by the one-loop approximation exactly (the
corresponding $\eps$ series terminate at first-order terms) and the
higher-order calculations for them are not needed.

However, the $\eps$ series for other important quantities do not terminate
and the calculation of the higher-order terms for them is of great interest.
In this paper, we present the results of the two-loop calculation for a
number of such quantities: the ultraviolet (UV) correction exponent $\omega$,
the Kolmogorov constant $C_{K}$ and the inertial-range skewness factor
${\cal S}$.

\section{The model, field theoretic formulation and renormalization}

Detailed exposition of the RG theory of turbulence and the bibliography
can be found in \cite{UFN,turbo}; below we restrict ourselves to only the
necessary information.

As the microscopic model of the fully developed, homogeneous, isotropic
turbulence of an incompressible viscous fluid one usually takes the
stochastic NS equation with a random driving force
\begin{equation}
\nabla _t\varphi _i=\nu _0\partial^{2} \varphi _i-\partial _i P+F_i ,
\qquad
\nabla _t\equiv \partial _t+(\varphi \partial)  .
\label{1.1}
\end{equation}
Here $\varphi _i$ is the transverse (due to the incompressibility)
three-dimensional vector
velocity field, $P$ and $F_i$ are the pressure and the transverse random
force per unit mass (all these quantities depend on $x\equiv\{t,{\bf x}\}$),
$\nu _0$ is the kinematical viscosity coefficient, $\partial^{2}$ is the
Laplace operator and $\nabla _t$ is the Lagrangian derivative. The problem
(\ref{1.1}) is studied on the entire $t$ axis and is augmented by the
retardation condition and the condition that $\varphi _i$ vanishes for
$t\to-\infty$. We assume for $F$ a Gaussian distribution with zero mean and
correlator
\begin{equation}
\langle F_i(x)F_j(x')\rangle =\delta (t-t')(2\pi )^{-3}\int
d{\bf k}\, P_{ij}({\bf k}) d_F(k)\exp \big[ {\rm i} {\bf k}
\left({\bf x}-{\bf x}'\right)\big]  ,
\label{1.2}
\end{equation}
where $P_{ij}({\bf k}) =\delta _{ij}  - k_i k_j / k^2$ is the transverse
projector and $d_F(k)$ is some function of $k\equiv |{\bf k}|$ and model
parameters.

The stochastic problem (\ref{1.1}), (\ref{1.2}) is equivalent to the field
theoretic model of the doubled set of fields $\Phi\equiv\{\varphi,\varphi'\}$
with action functional
\begin{equation}
S(\Phi )=\varphi 'D_F\varphi '/2+\varphi '[-\partial _t\varphi +\nu
_0\partial^{2} \varphi -(\varphi \partial )\varphi ] ,
\label{action}
\end{equation}
where $D_F$ is the random force correlator (\ref{1.2}) and the required
integrations over $x=\{t,{\bf x}\}$ and summations over the vector indices
are understood.

The standard RG technique can be applied to the model (\ref{action}) if
the function $d_{F}(k)$ is chosen in the form
\begin{equation}
d_F(k)=D_0\,k^{1-2\varepsilon}\,h(m/k), \qquad  h(0)=1.
\label{1.9}
\end{equation}
Here $D_0$ is the positive amplitude factor, $m=1/L$ is the reciprocal of the
integral turbulence scale $L$, the function $h(m/k)$ provides the IR
regularization and the exponent $\varepsilon>0$ plays the part of the RG
expansion parameter, similar to that played by $\eps=4-d$ in Wilson's theory
of critical phenomena \cite{Zinn}. The real (physical) value of this
parameter is $\varepsilon=2$: idealized energy injection by infinitely large
eddies corresponds to $d_{F}(k) \propto \delta({\bf k})$, and the function
(\ref{1.9}) for $\varepsilon\to2$ and the appropriate choice of the
amplitude can be considered as a power-law model of the three-dimensional
$\delta$ function.

The model (\ref{action}) is logarithmic (the coupling constant
$g_0\equiv D_0/\nu _0^3$ is dimensionless) at $\eps=0$, and the UV
divergences have the form of the
poles in $\eps$ in the correlation functions of the fields $\varphi$ and
$\varphi'$. Superficial UV divergences, whose removal requires counterterms,
are present only in the 1-irreducible function
$\langle\varphi'\varphi\rangle$, and the corresponding
counterterm has the form $\varphi'\partial^{2}\varphi$.
Thus for the complete elimination of the UV divergences
it is sufficient to perform the multiplicative renormalization of the
parameters $\nu_0$ and $g_{0}=D_{0}/\nu_0^{3}$
with the only independent renormalization constant $Z_{\nu}$:
\begin{equation}
\nu_0=\nu Z_{\nu}, \qquad g_{0}=g\mu^{2\eps}Z_{g},
\qquad Z_{g}=Z_{\nu}^{-3} \qquad (D_{0} = g_{0}\nu_0^{3}
= g\mu^{2\eps} \nu^{3}).
\label{18}
\end{equation}
Here $\mu$ is the reference mass in the minimal subtraction (MS) scheme,
which we always use in what follows, $g$ and $\nu$ are renormalized analogues
of the bare parameters $g_{0}$ and $\nu_0$, and $Z=Z(g,\eps,d)$ are the
renormalization constants. In the MS scheme they have the form ``1 +
only poles in $\eps$,'' in particular,
\begin{eqnarray}
Z_{\nu}=1+\sum _{k=1}^{\infty }a_k(g)\eps ^{-k}=1+\sum _{n=1}^{\infty }g^n
\sum _{k=1}^{n}a_{nk}\eps ^{-k} ,
\label{1.30}
\end{eqnarray}
with the one-loop coefficient $a_{11}=-1/40\pi^2$ \cite{Frisch,Pismak}.

\section{Two-loop approximation for the RG functions, fixed point and
the UV correction exponent}

We have performed calculation of the constant $Z_{\nu}$ with the accuracy of
$O(g^{2})$ (two-loop approximation). The calculation is rather involved and
will be presented elsewhere (some details can be found in \cite{twoloops}),
and below we give only the results for the residues
$a_{22}$ and $a_{21}$ at the second-order and first-order poles in $\eps$
in the representation (\ref{1.30}):
\begin{equation}
a_{22}/a_{11}^{2}= 1, \quad a_{21}/a_{11}^{2}\simeq - 1.65.
\label{AAA}
\end{equation}

The knowledge of the renormalization constant $Z_{\nu}$ to order $O(g^{2})$
allows for the calculation of the RG functions, the anomalous dimension
$\gamma_{\nu}$ and the beta function $\beta_{g}(\eps,g)$, with the
following accuracy:
\[\beta(g,\varepsilon) \equiv \Dm g =
g\left(-2\varepsilon+3\gamma_\nu(g)\right),\]
\begin{equation}
\gamma_\nu(g) \equiv \Dm \ln Z_{\nu} =
-2g\partial_g a_1(g)=-2\left(a_{11}g+2a_{21}g^2\right) +O(g^3),
\label{RGF}
\end{equation}
where $\Dm$ is the operation $\mu \partial/ \partial\mu$ at fixed bare
parameters. In the MS scheme only the residues at the first-order poles
in $\eps$, that is, only the coefficients $a_{k1}$, contribute to the RG
functions owing to the UV finiteness of the latter.

The coordinate of the fixed point is determined by the condition that
$\beta(g_{*})=0$. From (\ref{AAA}) and (\ref{RGF}) we thus obtain:
\begin{equation}
g_{*} = (40 \pi^2\eps/3) (1+\lambda\eps) + O(\eps^{3}),
\quad
\lambda\equiv 2a_{21}/3a_{11}^{2} \simeq -1.10.
\label{FP}
\end{equation}
The correction exponent $\omega$ is determined by the slope of the beta
function at the fixed point,  $\omega=\beta'(g_{*})$. Thus from (\ref{RGF})
and (\ref{FP}) we obtain the first and second terms of its $\eps$ expansion:
\begin{equation}
\omega=2\varepsilon(1-\lambda\varepsilon)+O(\varepsilon^3).
\label{Omega}
\end{equation}

\section{Two-loop calculation of the Kolmogorov constant}

The Kolmogorov constant $C_{K}$ can be defined as the (dimensionless)
coefficient in the inertial-range expression $S_{2}(r)=C_{K}(\E\,r)^{2/3}$
for the second-order structure function, predicted by the Kolmogorov--Obukhov
theory and confirmed by experiment \cite{Monin,Legacy}. Here $\E$ is the mean
energy dissipation rate and the $n$-th order (longitudinal, equal-time)
structure function is defined as
\begin{equation}
S_{n} (r) \equiv \big\langle [ \varphi_{r} (t,{\bf x}+{\bf r})
- \varphi_{r} (t,{\bf x})]^{n} \big\rangle, \qquad
\varphi_{r}\equiv (\varphi_{i}\cdot r_{i})/r, \quad r\equiv |{\bf r}|.
\label{struc}
\end{equation}
Using the exact relation $S_{3}(r)=-4\E\,r/5$ that follows from the energy
balance equation \cite{Monin,Legacy}, the constant $C_{K}$ can be related
to the inertial-range skewness factor:
\[ {\cal S} \equiv S_{3}/S_{2}^{3/2} = -(4/5)\,C_{K}^{-3/2}.\]

Many studies have been devoted to the derivation of $C_{K}$ within the
framework of the RG approach; see Refs. [11--18]. In order to
obtain $C_{K}$, it is necessary to augment the solution of the RG equation
for $S_{2}$ by some formula that relates the amplitude $D_{0}$ in the random
force correlator (\ref{1.9}) to the physical parameter $\E$. In particular,
in \cite{48,50} the first-order term of the $\eps$ expansion for the
pair correlator was combined with the so-called eddy damped quasi-Markovian
approximation for the energy transfer function, taken at $\eps=2$. More
elementary derivation, based on the exact relation between $\E$ and the
function $d_{F}(k)$ from (\ref{1.9}) was given in \cite{JETP}; see also
\cite{UFN,turbo}. In spite of the reasonable agreement with the experiment,
such derivations are not immaculate from the theoretical viewpoints. Their
common flaw is that the relation between $\E$ and $D_{0}$ is unambiguous
only in the limit $\eps\to2$:
\begin{equation}
\lim_{\eps\to2}\,\frac{D_0}{4\pi^2 (2-\eps)}=\E,
\label{WW}
\end{equation}
so that the coefficients of the corresponding $\eps$ expansions appear in
fact arbitrary; see the discussion in \cite{85} and Sec.~2.10
of~\cite{turbo}. The ambiguity is a consequence of the fact that the notion
of the Kolmogorov constant has no definite extension to the nonphysical
range $\eps<2$.

The experience on the RG theory of critical behaviour shows that unambiguous
$\eps$ expansions can be written for universal quantities, such as critical
exponents, normalized scaling functions and ratios of amplitudes in scaling
laws \cite{Zinn}. The constant $C_{K}$ extended to the range $\eps<2$ as in
\cite{48,50,JETP} involves a bare parameter, $D_{0}$, and
hence is not universal.

To circumvent this difficulty, we propose below an alternative derivation
that relates $C_{K}$ to an universal quantity and thus does not involve any
relation between $D_{0}$ and $\E$, and calculate $C_{K}$ to second order of
the expansion in $\eps$ (previous attempts have been confined with the first
order). Consider the ratio
\begin{equation}
Q(\eps)\equiv r\partial_{r} S_{2}(r) / |S_{3}(r)|^{2/3}=
r\partial_{r} S_{2}(r) / (-S_{3}(r))^{2/3}.
\label{RR}
\end{equation}
The operation $r\partial_{r}\equiv r\partial/\partial r$ kills the constant
contribution $\langle \varphi^{2} \rangle$ in $S_{2}$ that diverges as
$\Lambda\to\infty$ for $\eps<3/2$; in $S_{3}$ such constant contributions
are absent.

Solving the RG equations for the quantities in $Q(\eps)$ in the IR range
($\Lambda r \gg 1$) for general $0<\eps\le 2$ gives
\begin{equation}
S_{3}(r) = D_{0} r^{-3\Delta_{\varphi}} f_{3} (\eps), \quad
r\partial_{r} S_{2}(r) = D_{0}^{2/3} r^{-2\Delta_{\varphi}/3}
f_{2}(\eps), \quad \Delta_{\varphi} = 1-2\eps/3;
\label{RGS}
\end{equation}
see e.g. \cite{UFN,turbo}. Thus the quantity $Q(\eps)=f_{2}/(-f_{3})^{2/3}$
in (\ref{RR}) does not depend on
$D_{0}$ and can be calculated in the form of a regular $\eps$ expansion.
We calculated the scaling functions $f_{2,3}$ to the second order of the
$\eps$ expansion, which corresponds to the two-loop approximation in
(\ref{FP}), (\ref{Omega}), and obtained:
\begin{equation}
Q(\eps) = (1/3)(20\eps)^{1/3}\left[1+0.525\eps+O(\eps^2)\right].
\label{QQ}
\end{equation}
It is worth noting that the $\eps$ expansion for $f_{3}$ can be obtained
not only from the direct perturbative calculation, but also from the exact
expression
\begin{equation}
S_3(r)= -\frac{6 \Gamma(2-\eps)}
{2^{2\eps}\pi^{3/2}\Gamma(3/2+\eps)} \, D_{0}r^{-3\Delta_{\varphi}}
\label{Exa}
\end{equation}
that follows from the energy balance equation and in the limit $\eps\to2$,
along with the formula (\ref{WW}), reproduces the correct coefficient $-4/5$
(see above).

The value of $Q(\eps)$ at $\eps=2$ determines the Kolmogorov constant and
skewness factor through the exact relations $C_{K}=6\cdot 10^{-2/3}\,Q(2)$,
${\cal S}= - [1.5 \cdot Q(2)]^{-3/2}$, which follow from the definitions
and the identity $r\partial_{r}r^{\lambda}=\lambda r^{\lambda}$ for any
$\lambda$. Substituting the value of (\ref{QQ}) we obtain $C_{K}=3.02$,
${\cal S}= - 0.15$. If we retained only the first-order term in
(\ref{QQ}) we would have obtained $C_{K}=1.47$ and ${\cal S}= -0.45$.
We also recall the experimental estimates recommended in \cite{Monin}:
$C_{K} \approx 1.9$ and ${\cal S} \approx -0.28$.

\section{Conclusion}

We have accomplished the complete two-loop calculation of the renormalization
constant and RG functions for the stochastic problem (\ref{1.1})--(\ref{1.9})
and derived the coordinate of the fixed point, the UV correction exponent
$\omega$, the Kolmogorov constant $C_{K}$ and the inertial-range skewness
factor ${\cal S}$ to the second order of the corresponding $\eps$ expansions.
The new point is not
only the inclusion of the second-order correction, but also the derivation
of $C_{K}$ through an universal (in the sense of the theory of critical
behaviour) quantity.

Of course, one should have not expected that the second-order terms of the
$\eps$ expansions would be small in comparison to the first-order terms.
The experience on the RG theory of critical behaviour shows that such
corrections are not small for dynamical models (in contrast to static ones)
and for amplitudes (in contrast to exponents); see \cite{Zinn}. It is
thus rather surprising that in our case the account of the two-loop
contributions leads to reasonable changes in the results.

Although the $\eps^2$ correction to $\omega$ in (\ref{Omega}) is rather
large, it does not change its sign and hence does not destroy the IR
stability of the fixed point.

The first-order approximation $C_{K}=1.47$ underestimates, and the
second-order approximation $C_{K}=3.02$ overestimates the
co\-n\-ve\-n\-ti\-onal experimental value of the Kolmogorov constant
$C_{K} \approx 1.9$ \cite{Monin}.
Thus the experimental value of $C_{K}$ (and hence for ${\cal S}$) lies
in between of the two consecutive approximations. A similar situation is
encountered for the well-known Heisenberg model \cite{Monin}, where the
analogue of the Kolmogorov constant is known exactly and lies between
the first-order and second-order approximations given by the corresponding
$\eps$ expansion \cite{Hei}. If we assume, by the analogy with the
Heisenberg model, that the (unknown) exact predictions for $C_{K}$ and
${\cal S}$ lie between the first two approximations, we may conclude
that our calculation has given a very good estimate for these quantities.

The work was supported by the Nordic Grant for Network Cooperation with the
Baltic Countries and Northwest Russia No.~FIN-18/2001 and the GRACENAS Grant
No.~E00-3-24.

\end{document}